# Energy-Efficiency and Sustainability in New Generation Cloud Computing: A Vision and Directions for Integrated Management of Data Centre Resources and Workloads

Rajkumar Buyya[1], Shashikant Ilager[1], and Patricia Arroba[1,2]

[1] Cloud Computing and Distributed Systems (CLOUDS) Lab
School of Computing and Information Systems
The University of Melbourne, Australia
rbuyya@unimelb.edu.au, shashi.ilager@unimelb.edu.au

[2]ETSI Telecomunicación, CCS–Center for Computational Simulation
Universidad Politécnica de Madrid
Avenida Complutense, 30 Madrid 28040, Spain
p.arroba@upm.es

**Correspondence:** Rajkumar Buyya, Cloud Computing and Distributed Systems (CLOUDS) Lab
School of Computing and Information Systems. The University of Melbourne, Australia
Email: rbuyya@unimelb.edu.au

**Abstract:** Cloud computing has become a critical infrastructure for modern society, like electric power grids and roads. As the backbone of the modern economy, it offers subscription-based computing services anytime, anywhere, on a pay-as-you-go basis. Its use is growing exponentially with the continued development of new classes of applications driven by a huge number of emerging networked devices. However, the success of Cloud computing has created a new global energy challenge, as it comes at the cost of vast energy usage. Currently, data centres hosting Cloud services world-wide consume more energy than most countries. Globally, by 2025, they are projected to consume 20% of global electricity and emit up to 5.5% of the world's carbon emissions. In addition, a significant part of the energy consumed is transformed into heat which leads to operational problems, including a reduction in system reliability and the life expectancy of devices, and escalation in cooling requirements. Therefore, for future generations of Cloud computing to address the environmental and operational consequences of such significant energy usage, they must become energy-efficient and environmentally sustainable while continuing to deliver high-quality services.

In this paper, we propose a vision for learning-centric approach for the integrated management of new generation Cloud computing environments to reduce their energy consumption and carbon footprint while delivering service quality guarantees. In this paper, we identify the dimensions and key issues of integrated resource management and our envisioned approaches to address them. We present a conceptual architecture for energy-efficient new generation Clouds and early results on the integrated management of resources and workloads that evidence its potential benefits towards energy efficiency and sustainability.

## 1. Introduction

Cloud computing is pervasive in our economy and society as it hosts a wide variety of applications including enterprise, banking, transport, and healthcare, and delivers their services to consumers. Applications in the Cloud perform are significantly faster for users as they tap into the on-demand processing power of multiple servers on the backend. Further, to support emerging applications driven by IoT, Big Data, and Artificial Intelligence (AI), major Cloud service providers such as Amazon and Microsoft are deploying Data Centres (DCs) worldwide.

However, Cloud computing is causing an enormous escalation of energy usage, with increasingly pervasive front-end client devices, such as smartphones, interacting with DCs' back-end. A typical DC consumes as much energy as 25,000 households [6]. According to Greenpeace, DCs world-wide consume more energy than most countries and only the four largest economies (USA, China, Russia, Japan) surpass them in their annual electricity usage. Furthermore, there are over 8 million active DCs world-wide [8] and they are projected to grow over 12% per year. As their numbers and size grow, they will be consuming an increasing amount of energy.

Due to excessive reliance on brown energy (fossil fuel-based), DCs significantly contribute to the greenhouse gas emissions resulting in high carbon footprints (43 million tons of $CO_2$/year growing at an annual rate of 11%), and they are projected to contribute up to 5.5% of the world's carbon emissions by 2025 [2], and by 2030, they would consume about 8000 TWh (see Figure 1) [3]. Thus, it is imperative to have strategies for DCs to reduce their carbon footprints in our carbon-constrained future. Although harnessing renewable energy addresses this issue, its availability can be intermittent, and energy efficiency within data centres also continues to remain a significant problem to be solved. Moreover, 93% of the energy used in DCs support ICT equipment operation and cooling [17] (see Table 1). Therefore, we envision the reduction of the energy needs and carbon footprint substantially through the integrated management of all resources.

**Table 1:** Data Centre Energy Usage by its Different Components.

| DC Components | %Use |
|---|---|
| Servers | 50% |
| Cooling Systems | 35% |
| Networks | 8% |
| Others (UPS,PDU, etc.) | 7% |

Through this research, with focus on integrated energy efficient management of DC resources, we aim to bring down the energy consumption of DCs world-wide up to 80%, from 8000 TWh (worst case) in 2030 to about 1200 TWh (see Figure 1) [3]. Therefore, there is a need for a new approach for the management of DCs, where every component is instrumented and dealt in an integrated manner to ensure the energy-efficient execution of applications [12].

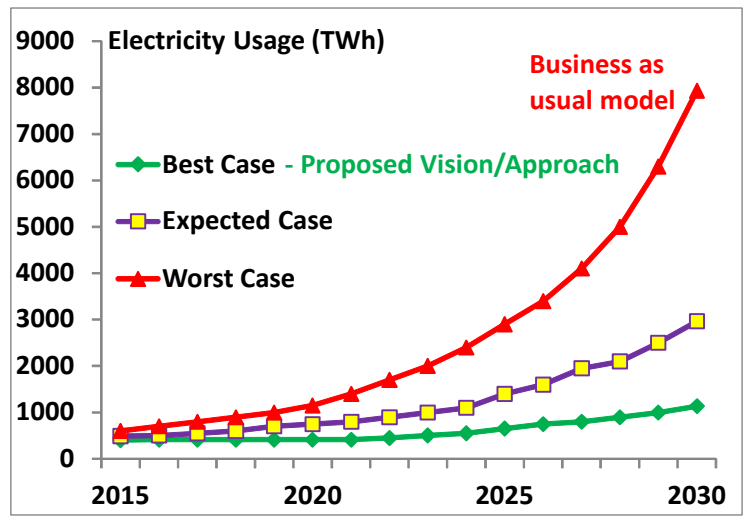

**Figure 1:** Electricity Use by Data Centres Worldwide.

With the emergence of pervasive IoT-based applications, demand for low-latency (i.e., near real-time) services has increased tremendously [36] and is creating a highly distributed environment (see Figure 2). Applications such as smart cars, smart cities, and smart healthcare require edge-affinity to the DC services to meet their low latency Quality of Service (QoS) requirements. Such developments will lead to the deployment of many DCs at the edge and across many regions with different capabilities and energy sources. Monitoring Agents (MA) monitors state of the data centre resources and applications and continuously provides live data to and help to take the informed decisions for optimizing data centre resources. Brokering Agents (BAs) as part of a DC resource management seamlessly offload tasks to other DCs. By dynamically offloading requests, they facilitate harnessing green energy, edge affinity, and exploiting the availability of a smart grid's energy at a low price. Thus, the proposed research approach enables the optimal use of resources, network traffic reduction, service cost reduction, and increased operational efficiency, along with the reduction in DCs' energy consumption and carbon footprint, while delivering Cloud services that meet the QoS requirements of end users.

In this direction, this paper introduces our vision towards energy efficiency and sustainability in the new generation of DCs, including: (i) minimise energy consumption through sustainable use of resources while maintaining QoS demand of users, (ii) investigate energy efficiency across all types of DC resources, application models and Cloud services, and (iii) achieve a paradigm shift in management systems from control-centric to dynamic learning-centric algorithms.

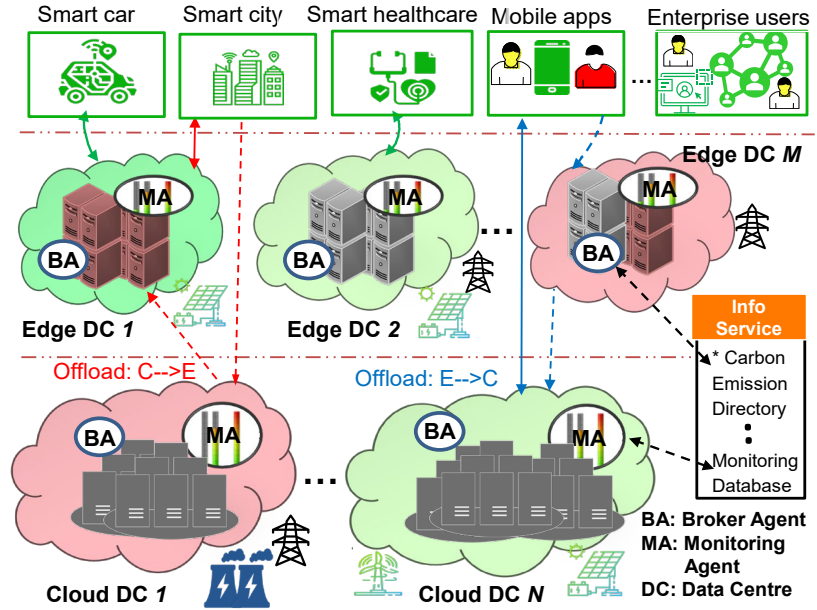

**Figure 2:** A Distributed Environment for Enterprise and Smart * Applications with Edge and Cloud Data Centres.

The rest of the paper is organised as follows. Section 2 introduces the significance and innovations of our energy-centric proposal. The dimensions of integrated management as well as the research questions that arise from them are discussed in Section 3. In Section 4, we present our conceptual architecture for energy-efficient new generation Cloud computing. Section 5 identifies the key issues of integrated resource management systems and our envisioned approaches to address them. A case study for critical-business applications is presented in Section 6 to demonstrate the potential impact of the integrated resource management on energy efficiency. In Section 7, conclusions and guidelines for future research are drawn.

## 2. Significance and Innovation of Proposed Vision

In the state-of-the-art, systematic alignment for energy efficiency across all DC resources and service models with integrated management has not been prioritised to date. Existing approaches are confined to energy efficiency at a "single" level of the Cloud computing stack [11], i.e., either at the infrastructure or platform level. In addition, they primarily focus on a "single" type of resource (e.g., server only) independently, without explicitly considering other types (e.g., cooling systems) [5] and can achieve only up to 25% reduction in energy use [13]. The same approach is followed while dealing with users and their application workloads. Such approaches are limited in their effectiveness because of the complexity and interrelationship between different elements of the Cloud computing stack and resources. They also lead to unintended consequences such as the creation of hot spots, which can cause system failures and escalate cooling energy requirements. Therefore, to achieve significant reductions in energy consumption, in the range of 50% to 80%, a quantum leap in research is required. We propose to go beyond the current research boundaries to enable integrated management of major power consuming elements of DCs – servers, cooling systems, and networks – to enhance energy efficiency, resilience, and sustainability.

The innovations presented in this paper will fundamentally change the way energy efficiency and sustainability are aligned, engineered, and realised in new generation Clouds. Through close synthesis of theory and practice, it will deliver timely innovations for significant scientific, economic, environmental, and societal impacts.

## 2.1 Delivering Service Quality Guarantees

Clouds hosting time-critical and always-on applications need QoS guarantees such as availability, response time/deadline [28], cost, and throughput along with tolerating faults. We propose approaches that addresses service quality guarantees through the development of QoS-driven provisioning and scheduling of resources along with energy-efficient techniques for execution of applications. We envision future DCs that are energy-efficient, resilient, available 24/7, and that have a minimal impact on the environment at a time when the huge growth in Cloud computing is at odds with the world's need for a smaller carbon footprint. Achieving an efficient trade-off between the conflicting objectives of optimising energy efficiency and minimising execution time/cost is challenging due to the inherent diversity associated with Cloud workloads, and heterogeneity of underlying infrastructures. By harnessing geographically distributed DCs, we can route requests with edge affinity and deliver time-critical and reliable services to users in a cost-efficient manner. Thus, the resulting scientific innovations are both timely and extremely important to provide a fault-tolerant scenario for new Cloud infrastructures.

## 2.2 Integrated Resource Management for Energy Efficiency

One of the main sources of energy inefficiencies in DCs is that its servers are often underutilised, only using about 10 to 50% of their capacity [18]. This problem is amplified by the fact that their energy consumption is not in proportion to their load (idle servers often consume up to 30 % of energy equivalent to their full load in virtualized Cloud environments [34]). Consolidation of Virtual Machines (VMs) is used to reduce the number of active servers and increase resource utilisation to minimise server energy usage. However, this approach typically ignores other core components of high energy use, such as networks and cooling systems. It also potentially increases the number of hot spots in DCs, increasing the likelihood of server failure besides increasing cooling requirements. On the other hand, hotspot mitigation can be performed by spreading load across servers and racks, but it will place communicating VMs far apart, which increases network latency and deteriorates service quality. In this context, we would require novel integrated resource management algorithms that considers all these elements and manages trade-offs. Through the interplay between user QoS requirements, workloads, energy sources, and IoT-enabled cooling systems, we can dynamically manage resources in time and space dimensions.

## 2.3 New Learning-based Algorithms for Cloud Resource Management Systems (RMS)

DCs are enormously complex, and it is hard to manually fine-tune their parameters for energy efficiency and performance. This is due to the large number of heterogeneous servers hosting a wide variety of applications and complex interactions between servers, networks, and cooling systems. For example, just 10 pieces of equipment, each with 10 settings, would have $10^{10}$ possible configurations, a set of possibilities impossible to manually fine-tune. Current heuristics/rule-based solutions are unable to provide high efficiency as they are static, manually fine-tuned and cannot adjust to the complex dynamic contexts. We propose Machine Learning (ML) and Reinforcement Learning (RL)-based methods to employ proactive measures and adapt to the complex dynamicity to deliver energy efficiency across all resources and the whole system. These algorithms will be most efficient as they dynamically learn the complexity of Cloud environments by incorporating the experience, the current status and the workload characteristics.

These innovations embody a set of highly novel theories, principles, and techniques that together will produce new learning-centric algorithms and systems for the integrated management of new generation Cloud computing environments to reduce their energy consumption and carbon footprint.

## 3. Dimensions and Research Challenges

Achieving sustainable and energy-efficient new generation Clouds is a complex challenge as it impacts from many dimensions. Computing and cooling resources, application models, nature of workloads, deployment models, energy sources, monitoring systems and user QoS demand are dimensions of integrated management identified in our research as shown in Figure 3. Integrated management involves considering every aspect of all the elements and their effect on other domains, making learning-centric approaches an interesting option to explore. This complex and challenging cloud computing environment raises the following key research questions:

- How to trade-off QoS requirements of different users?
- Which is the most optimal physical machine for hosting a VM for a specific application and network model?
- When to provision or de-provision resources for a given workload?

- How to consolidate virtual resources (which, when, where) to reduce energy consumption without compromising QoS?
- What is the best operating frequency of computing resources to minimise energy and meet QoS?
- How to manage/schedule workloads without creating hotspots while meeting the QoS of the applications?
- How to accurately predict/estimate the thermal status of the Cloud data centre and the temperature of its resources?
- What is the best temperature-aware approach for the placement of workloads to avoid failures/outages and minimise cooling energy (reducing peak temperature).
- How to harness temperature sensors data to reduce cooling systems' energy consumption by varying the supply air temperature?
- How to organise workloads and harness diverse variable renewable energy sources (wind and solar power among others) to minimise carbon footprint?
- How to balance the load across geo-distributed data centres to ensure sustainability?

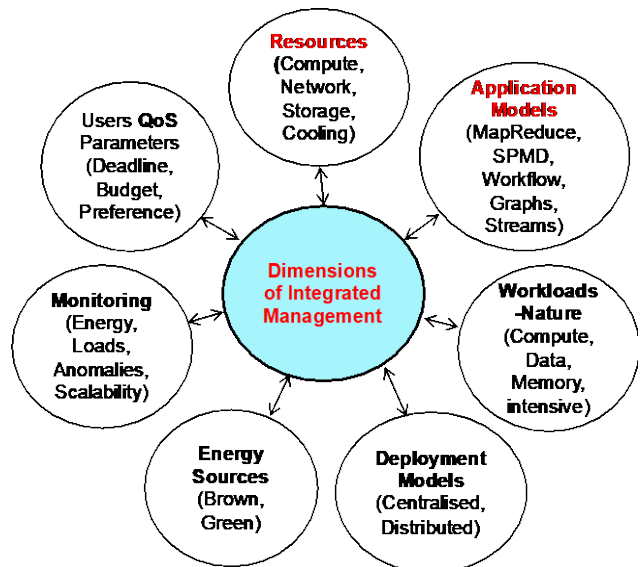

**Figure 3**: Dimensions of Integrated Management

## 4. Architectural Framework

The aim of our vision is to consider all aspects of energy efficiency and sustainability in new generation Cloud computing across architectures, algorithms, software systems, and applications, dealing with device and system-wide resilience, energy and thermal issues. To this end, our approach is driven by a set of key requirements: (1) energy efficiency across all types of DC resources, Cloud services and application models, (2) shift from control-centric to dynamic learning centric algorithms in system management, and (3) sustainable use of resources of DCs to minimise energy consumption and carbon footprint and meet QoS demands of users. They can be met through integrated management of resources, interplay between different service layers, and use of machine-learning based methods for proactive management of resources.

Figure 4 illustrates the high-level architectural framework guiding our approach and the main elements of the **new energy-efficient, integrated resource management system**. From the top, service consumers define their QoS requirements and access Cloud services, submitting their requests from anywhere in the world. Our proposed RMS is responsible for energy-efficient management of user requests, applications, and all resources of DCs, which can be Edge Data Centres (EDCs) or remote Cloud Data Centres (CDCs) to deliver quality services. The RMS can only achieve its goals if algorithms and techniques for energy-efficient Cloud computing are integrated at all levels of the Cloud stack, from the physical layer (comprising DC's computing, networking, and cooling resources) to the application layer. To this end, the DC must contain heat sensors that monitor the temperature at different places, power sensors that monitor energy consumption of servers and networking equipment, and programmable cooling systems.

Our RMS implements the logic enabling fine grained control of cooling systems and load distribution across servers to avoid hotspots. To ensure the QoS of accepted requests, the entry point of the RMS will be an admission control module that filters out requests that cannot be served within the desired QoS. RMS takes care of provisioning of resources and scheduling application tasks to suitable resources. As regular monitoring is critical in decision making, RMS closely monitors and manages the health and load of the servers and the status of the cooling system to proactively balance the overall system for energy efficiency. A green resource broker agent handles the distribution of requests across multiple DCs to reduce their carbon footprint and meet user QoS requirements.

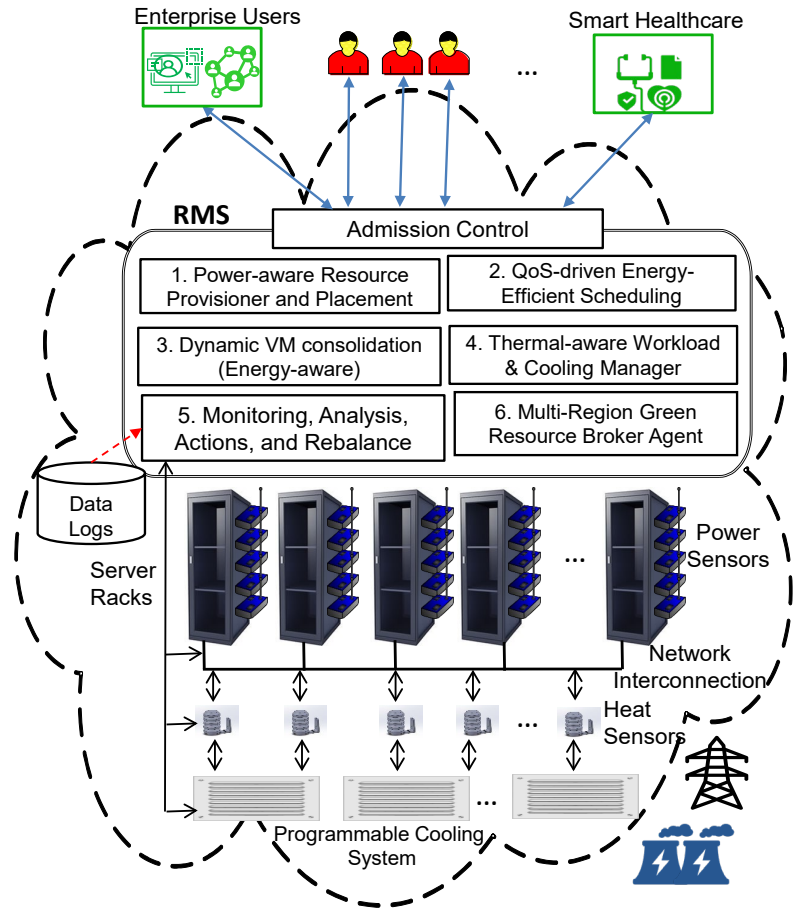

**Figure 4**: A data centre resource management system.

## 5. Research Issues and Envisioned Approaches as Future Directions

As we aim towards the development of our new integrated resource management system, we identified some issues in essential areas related to energy efficiency that need to be addressed. In this section, we present the key issues and the approaches we envision to solve them.

### 5.1 Workload and Power-aware Dynamic Resource Provisioning and Placement

Clouds allocate resources to applications dynamically based on users' requirements. These resources are offered as VMs or containers with various configurations ready to be leased and used as long as they are needed. This model calls for a resource provisioning strategy that works together with the scheduling algorithm driven by QoS requirements. Provisioning of virtual resources for user requests includes estimating the number of required resources and providing suitable resources accordingly. A broad range of applications executed in Clouds differ from each other in terms of duration (ranging from web transactions running for a fraction of a second to streaming applications that run continuously), use of resources (e.g., CPU or I/O-intensive), degree of parallelism, and dependencies between tasks (which dictate the maximum number of tasks that can be executed concurrently). The huge variation in application characteristics means that no single provisioning solution can optimise and support all of them. When the same provisioning approach is utilised for different applications, they exhibit a huge difference in performance [9].

To counter this limitation of "one size fits all" provisioning, we aim to develop provisioning algorithms that are application structure-aware and balance both QoS and energy utilisation. Provisioning of virtual machines and containers raises another infrastructure issue, namely the *placement of VMs*. This problem includes the selection of the best suited physical machine (PM), i.e., server, for each VM, taking into consideration requirements of the VM, current usage and energy utilisation levels of available servers in the DC, and the structure of applications to maintain the VMs-affinity for their interrelated communicating tasks. We aim to address this through placement policies to maintain meta-data of VMs containing information about application topologies.

**5.2 QoS-driven Energy-Efficient Scheduling of Applications**
In creating elastic/scalable applications for execution on Clouds, a variety of application programming models, such as task farming, workflow, and stream-oriented models are used. Each of these models needs its own scheduling algorithm. Most of the prior works [23] [38] have focused on the general problem of energy-efficient scheduling without focusing on fine tuning the scheduling mechanisms to suit workload characteristics and all types of resources. To illustrate our learning-based scheduling approach, let's consider a scheduling model for workflow-oriented applications, which are common in scientific and business domains. In workflow applications, precedence constraints among tasks and corresponding inter-task communications are amongst the unique characteristics that need to be exploited to optimise the utilisation of resources. Each workflow task can be either batch or stream-based, and their performance can vary depending on the type of resources on which the tasks are scheduled.

As shown in Figure 5, our proposed Deep Reinforcement Learning (DRL)-based RMS interacts with the DC environment to learn intelligent scheduling policies. Its key components are: (1) ***scheduling agent***, which observes the current state of the system, takes an action (scheduling decision) and sends it to the application execution manager, and fetches a reward from the reward generator for the current action; (2) ***application execution manager***, which takes care of submission of tasks to the DC VMs and monitoring resource usage, resource availability, and system health; (3) ***reward generator***, which uses the monitoring system to generate RL reward that can be used to train RL-based scheduling agents.

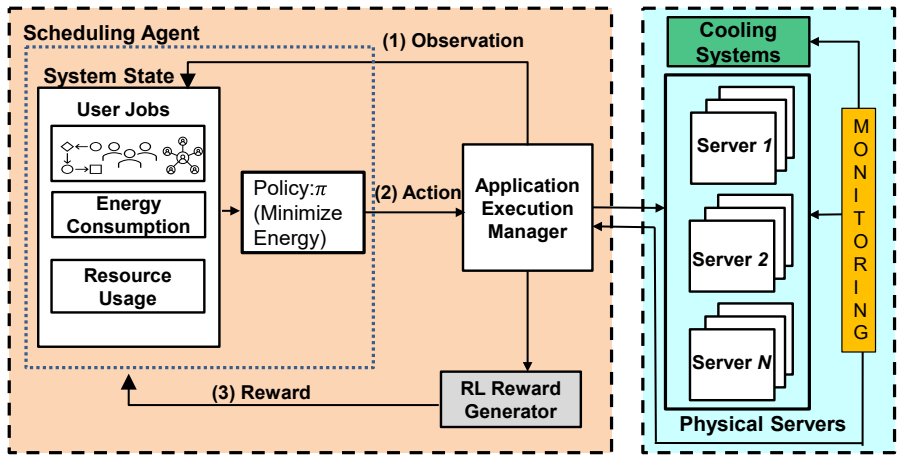

**Figure 5**: DRL-based Energy-Efficient Scheduling System

A common weakness of existing DRL-based techniques [24] is that their agents are provided with a limited representation of the actual environment and as a result, agents learn suboptimal resource allocation policies. In contrast, we will facilitate the agents to learn integrated resource allocation policies by formulating the state-spaces to incorporate the statuses of multiple components including servers, networks as well as cooling systems. The agents will learn to optimise the utilisation of resources and establish the right trade-off among multiple conflicting objectives by learning through their interactions with the operating environments. We will create advanced reward structures coupled with comprehensive state representations for expediting the learning process of the agents. The policies thus learnt will surpass the limitations associated with human designed heuristics and mathematical optimisations with high time complexities. Whenever provisioned resources are unable to meet QoS requirements, the scheduler requests additional resources from the resource provisioner and interacts with the power state manager to configure optimal frequencies to execute tasks with minimal energy consumption. To this end we aim to develop QoS-aware scheduling policies to enhance utilisation and energy efficiency of underlying DC infrastructures by leveraging the unique characteristics of application models.

### 5.3 Application-Aware Management of Power States of Computing Elements

The speed and power consumption of computing elements such as CPUs, GPUs, TPUs, and memory are controlled by regulating their clock frequencies and, accordingly, transitioning into different power states. However, their power management is a complex phenomenon due to nonlinear power curves of different computing elements for different applications. For instance, the execution time of compute-intensive tasks reduces linearly, whereas power consumption reduces quadratically with reduction in CPU frequency [15]. Hence, managing efficient power states becomes a significant challenge in such conditions. In this direction, our proposed approach is to develop mechanisms for management of power states of computing elements and schedule tasks in a power-aware manner to reduce the energy utilisation of DCs through DVFS (Dynamic Voltage and Frequency Scaling).

A memory-intensive application would require higher memory frequency compared to CPU frequencies, and such heterogeneity needs to be accounted for in the solution. In our recent work [21], we proposed data-driven ML techniques for an energy-efficient frequency scaling of GPUs for commonly used benchmarking applications. We created prediction models that learn different applications' behaviour on different frequency settings, which helps the scheduler to configure energy-efficient frequencies while meeting applications' QoS requirements. Building on our recent work, we propose to investigate how to dynamically configure operating frequencies and power states of different elements to execute a variety of applications with minimal energy consumption.

### 5.4 Energy-aware Dynamic Consolidation

Consolidation plays a vital role in minimising the DCs energy consumption as it focuses on how to minimise the number of active physical machines (PMs) hosting VMs so that inactive PMs can be shut down or placed into idle mode. This is an important requirement as servers consume a significant amount of energy even at their lower utilisation. Consolidation helps to utilise the active PMs efficiently by migrating VMs from underutilised PMs to suitable PMs. However, this should be done without introducing significant migration overheads and ensuring service quality of applications. Hence, we propose to go beyond simply consolidating users' applications based on what resources they currently require meeting their current QoS requirements. Users' activities and performance requirements both vary in time, in ways that may to some extent be predictable. We aim to answer the question: *Can the observed history of the application behaviour lead to better workload consolidation? That is, can measurement lead to reduced physical resources required, improved application performance, or reduced need for migration?* We propose to address this issue by developing consolidation algorithms that will use the information about the historical workload patterns and application behaviour to select which applications will share physical resources. These will minimise overlapping of the applications' resource usage, and thus their influence on each other, as well as reducing the amount of migration needed as applications' need change. A consolidation algorithm can be designed as a centralised or decentralized algorithm, depending on the infrastructure and its workload requirements. A consolidation algorithm needs to have knowledge of all the resources and workload status in data centre, in such case, a centralized controller is feasible where we can aggregate information and take the informed decisions accordingly [32]. However, this may not be feasible or efficient in multi-cloud scenarios, where data centres are geographically distributed and may incur latency and communication overhead. A decentralized consolidation algorithm, on the other hand, allows local communication and optimization among data centres [37], which can improve resource utilization and performance.

In multi-component applications, VMs' affinity needs to be maintained to reduce the network latency between their components. In addition, consolidation techniques should also avoid creating hotspots as overutilisation may lead to increased heat dissipation. Currently, these issues are addressed independently. In this regard, we aim to create integrated consolidation policies that reduce energy consumption and the number of migrations while meeting QoS requirements of applications.

### 5.5 Thermal-aware Management of Workloads and Cooling Systems

In DCs, IT resources such as servers dissipate an enormous amount of heat, rising CPU temperatures up to 100 $^0$C. This has a significant impact on the energy consumption of cooling systems. It also creates hot spots and triggers CPU thermal throttling mechanisms decreasing CPU speed, causing performance degradation of applications, and damage to silicon components of servers affecting the reliability of the system. These problems can be overcome by thermal-aware workload management solutions, but they require accurate server temperature estimation under varied workload conditions and managing the server load accordingly. Existing solutions rely on analytical or Computational Fluid Dynamics (CFD) based techniques to estimate the ambient temperature of servers. As they are compute-intensive and/or extremely sensitive to changes in the DC layout, they are not suitable for use in RMS making online/real-time decisions.

In this context, we envision the deployment of heat and humidity IoT-sensors, and to enable system load and power sensors on each server to monitor the environment and system activities. This data may be integrated with monitoring systems and analysed to identify hotspots in the DC caused by uneven distribution of load across servers. We propose to use the collected data and build data-driven machine learning models to predict server temperatures accurately. The runtime inference of such models will be accurate and faster compared to the existing solutions. Our approach involves the development of new thermal-management algorithms that monitor workloads and estimate the thermal state of physical nodes using ML models and reallocate workload from the overheated nodes to others. We aim to leverage temperature variations between different workloads and swap them at an appropriate time to control the temperature. Hardware level techniques, such as DVFS, can lower the temperature when it surpasses the thermal threshold. They can be effectively used whenever the QoS of hosted applications does not require processors to operate at full capacity. Along with workload management, we propose algorithms and models to aid cooling-systems to configure the optimal parameters, including the selection of suitable air temperature supply, and server fan speeds, for energy efficiency.

### 5.6 Monitoring, Analysis, and Actions for Rebalancing the System

Integrated management of DCs require continuous monitoring of their resources, analysis of the monitored data, and taking appropriate actions in runtime to increase efficiency. In this context our objective is to answer the question: *How the monitored data of various sensors can be used to better balancing of the data centre from an energy efficiency perspective?* RMSs are highly dependent on infrastructure monitoring systems as they provide insights on how well resources are delivering promised services. Moreover, learning-based algorithms such as ML techniques depend on the quality and quantity of the data used to train the models. In DCs, getting access to such data is challenging due to: (1) different resources (server, network elements), and cooling systems collect the data using different monitoring tools and sensors, (2) lack of common semantics and standards between them, and (3) the need for complex and tedious mechanisms to unify these data in a meaningful manner. Hence, we envision a monitoring system leveraging existing platforms that are agnostic to the underlying heterogeneity and combine different subsystems' data with common semantics. This integrated, monitored data will be fed to different RMS modules to analyse and take actions accordingly. For example, in overload conditions, brownout-based algorithms will be used to deactivate the optional components of applications for efficiency.

ML and RL algorithms face issues such as learning from limited data and cold-start problems. Specifically, RL algorithms consume large amounts of time to learn efficient policies and face real world challenges including: (1) safety constraints that should never or at least rarely be violated, (2) the need for reward functions that are unspecified, multi-objective, or risk-sensitive, and (3) inferences that must happen in real-time at the control frequency of the system [25]. To address these issues, we propose to create transfer learning and asynchronous multi-agent-based solutions and harness the monitored data to train the RL agents in simulation to improve and perform better over time before real world deployment. These algorithms will determine automated actions to rebalance the system. These include VM migration, load balancing, and adjusting the setting of room cooling systems and the fan speed of servers interacting with RMS modules.

### 5.7 Harnessing Multi-Region Edge and Cloud Data Centres with Varied Energy Sources

Cloud service providers have been deploying DCs in multiple regions to offer edge-oriented services especially targeting IoT-based applications to support their low-latency requirements (see Figure 2). These edge DCs are often powered using renewable energy at different times of a day depending on weather conditions. Hence, maximising renewable energy usage is crucial to reduce their carbon footprint. However, renewable energy is intermittent and uncontrollable due to dynamic weather conditions. This poses a significant challenge to Cloud workloads that run in the always-on mode. Although energy storage system such as uninterrupted power supply (UPS) or dedicated battery storage systems can deal with intermittent availability issue at a certain level [39], however such energy storage system can be expensive and lead to energy leakage. Thus, it may not fully solve the intermittent availability issue of renewable energy. To address this issue, we propose to explore workload offloading techniques across edge-cloud DCs to increase renewable energy usage and maintain edge-affinity to users for their latency sensitive IoT workloads. This reduces the use of brown energy, facilitates the widespread use of clean energy, and addresses the issue of intermittent availability. We aim to create prediction models to forecast the availability of renewable energy at different DCs, workload level in DCs, and admit the workloads into DCs accordingly. These policies may adopt job-delaying strategies in case of batch jobs, to the period where renewable energy usage can be maximised. It is also essential to investigate how the choice of physical network infrastructure and virtual network topology affect offloading costs (delay and energy) with the increased amount of data transfer.

Edge-cloud data centres deployed at multiple regions provide an opportunity to reduce energy-related costs. The local demand for electricity varies with time and weather conditions. This causes time-varying differences in the price of electricity [35] at each location, along with different environmental costs (due to different sources of energy – coal, wind, solar). This gives scope to adjust the load sent to each region to improve cost-efficiency. We envision algorithms for these offloading decisions that consider the location of the user, the power efficiency of resources at each site, the energy mix, taking advantage of renewable energy generation and managing energy storage according to its price variations [20], and the number of servers currently powered-on in each region. This last factor is important to avoid excess switching of servers on or off as loads shift. We also propose to investigate how this flexibility in electricity demand can allow Cloud operators to become active participants in the smart power grid.

## 6. Case Study: Energy-Efficient Clouds for Business-critical Applications

In this section, we present a use case of integrated resource management in the context of business-critical applications to demonstrate the impact of moving towards an energy-centric approach while maintaining the required QoS. Business-critical applications are workloads that must necessarily be operational for the proper functioning of companies. These applications involve front-end and back-end services and cover domains such as customer relationship management, email, databases, and financial modelling and simulation. If the performance of these workloads (web servers, mail servers, application servers, etc.) is compromised, it can cause businesses to lose customers, productivity, and revenue. Ensuring QoS for business-critical applications is therefore a major challenge in increasing the sustainability of new-generation Clouds.

However, we are facing a scenario full of challenges for performance-centric approaches. The resource utilisation of these applications is usually low but is very dynamic, with daily peaks exceeding the average CPU utilisation by up to 100 times [27]. To reduce power consumption in these high variability scenarios, typical aggressive VM consolidation policies can lead to two problems. On the one hand, VM oversubscription can lead to loss of QoS, which is unacceptable for this type of workload. On the other hand, undesirable hot spots can be generated in the data centre, increasing cooling consumption, and causing irreparable damage to equipment and service outages. From the above, we can note that energy-centric approaches that are conscious of the integrated consumption and temperature of both IT and cooling systems are strategic to provide reliable business-critical applications, while maximising the sustainability of the Clouds that execute them. In this context, our case study illustrates some aspects of our vision, as the application-aware management of power states of computing elements and the thermal-aware management of workloads and cooling systems.

### 6.1 Integrated thermal-aware resource management

Figure 6 shows the overview of the integrated resource management system in a Cloud DC. Integrated management requires monitoring of both cooling and computing systems. In this scenario, we included the setpoint temperature of the cooling system since its consumption and the thermal distribution of the data room depend on this parameter. The inlet temperature of the servers is provided by a network of sensors. This temperature is relevant as it has an impact on the temperature reached by the internal components of the servers such as CPUs. The monitoring system polls servers' onboard sensors from the racks to obtain the necessary data to sense the thermal and power distribution of the data room, and to train the models that should help us predict their behaviour. This monitoring system can feed both a digital twin, to provide the current state of the DC, as well as the ML-based models. Thus, the models learn about the thermal and power behaviour of the room to provide very accurate estimations.

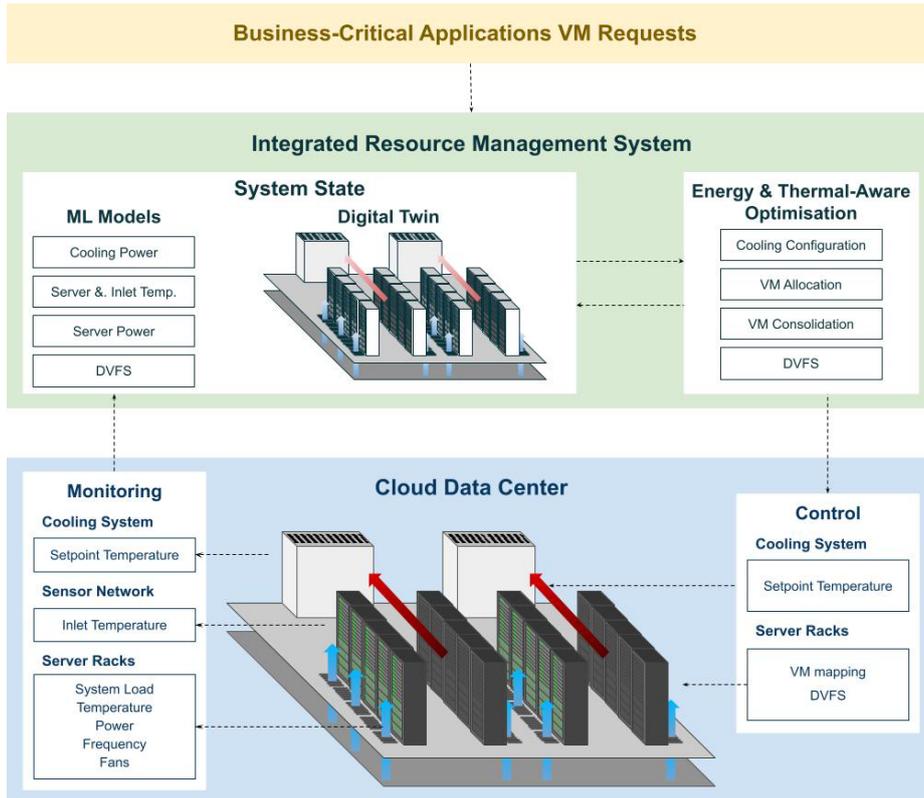

**Figure 6**: Overview of the energy-centric integrated resource management system.

The integrated resource management system uses the ML models to predict the impact of the energy and temperature-aware optimisation on the digital twin. In this way, the system knows in advance the performance of the VM allocation and consolidation approaches on the DVFS modes, the thermal distribution, the IT and cooling power consumption, and the QoS. Once the optimisation algorithm finds the best configuration scenario in the digital twin, it uses this setup to control the setpoint temperature of the cooling system, and both the mapping of VMs and the DVFS configuration of the server racks.

**6.2 Performance Evaluation**

We present the evaluation of the integrated resource management system of our energy-centric approach with the objective of improving the efficiency of the Cloud DC. To evaluate the performance of our approach, we use real traces from the DCs of Bitbrains, a company that provides Cloud computing IT services to customers in the banking and insurance sector. The workload involves running business-critical applications serving users in financial institutions and engineering firms [27].

The VMs run management servers for the day-to-day operation of user environments, application servers such as web and database servers, and compute nodes that perform intensive computing such as simulations. The traces include dynamic CPU, memory, disk, and network utilisation values. Bitbrains uses a high-performance policy on the servers that keeps CPUs at their maximum frequency, so these utilisation values always have the same frequency reference. To achieve a controllable environment, we have chosen the CloudSim [29] toolkit to simulate a Cloud computing DC. In this work, we have extended the simulator to include the cooling infrastructure and both the thermal behaviour and DVFS configuration of the servers. ML-based models have been used for modelling the cooling power consumption, internal and inlet temperature of servers, power consumption of servers, and for dynamic control of the DVFS modes of servers.

Our experimental setup includes a total of 1127 VMs, where we carry the optimization with interval of 300 seconds, i.e., resource demand assessment and resource allocation. We use four VM types from general flavors. Configurations of these VMs are: (a) 1 core, 4GB RAM, (b) 2 cores, 8GB RAM, and (c) 4 cores, 16GB RAM. The simulation scenario involves a Cloud DC consisting of 1200 hosts, specifically modelled as Fujitsu RX300 S6 servers equipped with an Intel Xeon E5620 Quad Core processor running at a maximum clock speed of 2.4 GHz. These servers have 16 GB of RAM, 1 GB of storage capacity, and run a 64-bit CentOS 6.4 operating system virtualized by the QEMU-KVM hypervisor. Individual server operating frequencies can be set at 1.73, 1.86, 2.13, 2.26, 2.39, and 2.40 GHz. Throughout the simulations, the active server count is notably decreased by enabling oversubscription. To avoid fan failures, the server inlet temperature is limited to a maximum of 303 K. Additional information on the workload, simulation configuration, and ML-based thermal and power models can be found in our previous research works [30][31].

Our energy and thermal-aware optimisation process is described as follows. This process supports energy efficient VM allocation for incoming requests and for reallocating existing VMs due to consolidation. First, the integrated resource management system detects overloaded and underloaded servers. Based on this, the system recommends the migration of certain VMs to reduce performance degradation, hotspot emergence, and to maximise resource utilisation.

Then, the system uses the ML-based models to predict the impact of allocating each VM on each of the servers with available resources, following a best fit decreasing approach iteratively. Given the outputs of the models (such as DVFS modes for servers, their temperature and utilization) all the proposed mappings for each VM are sorted using an optimisation criterion. The proposed criterion maximises the utilisation of servers with the minimum increase in frequency since there is a quadratic dependence of the energy consumption with the frequency. The resource management system updates the digital twin allocating each VM to the best server according to this criterion. It also uses the ML-based models to update the thermal and power status of the digital twin. The process is then repeated to assign the next VM until all the VMs have been assigned.

At the end of this optimisation process, the system delivers: (i) the proposed mapping of the entire set of VMs on the servers in the DC racks, (ii) the optimised DVFS configuration of each server to ensure performance, and (iii) the setpoint temperature of the cooling system to avoid equipment damage considering this mapping. Finally, the integrated resource management system provides the actions to control the actual cooling and computing infrastructures knobs including the optimal configuration of the frequency of each server, the air temperature supplied by the cooling system, and the VM allocation policy according to the observed impact on the digital twin.

We compare the results of our work with two state-of-the-art algorithms. The first aims to minimise the increase in server power consumption for the allocation of each VM with the objective of reducing IT consumption [32]. The second aims to maximise this power increment with the objective of increasing server rack utilisation [33]. Neither of these approaches considers temperature or DVFS for decision making so a fixed setpoint temperature of 291K is chosen to ensure that the servers are kept at a safe temperature.

The results show that our energy-centric approach that is aware of the impact of room temperature and DVFS can reduce on average the total consumption of the Cloud DC by 14%, presenting a 58% reduction in cooling power consumption. The PUE, compared to baselines, is reduced from 1.37 to a value of 1.16. Our strategy maintains the QoS with respect to baselines, avoiding server overload and guaranteeing dynamic operating frequencies that preserve the performance of high variability business-critical applications.

## 7. Conclusions and Final Remarks

In our digital world, Clouds have become the preferred computing platform for hosting business and government applications. Current Cloud computing services, however, offer execution of applications with little emphasis on making Clouds more energy-efficient and sustainable. This issue is going to be even more critical as DCs world-wide are projected to consume 20% of global electricity by 2025. We aim at solving this critical problem by achieving a quantum leap in energy efficiency and sustainability for next generation Cloud computing, dealing with energy efficiency through integrated management of all resources and QoS requirements of applications.

This paper presents our vision towards energy-efficient new generation Cloud computing. Next generation Clouds should be able to reduce their energy consumption and carbon footprint, by creating new learning-centric algorithms and software systems for integrated management of all their related resources and applications. Our early experiments show the potential of our energy-centric approach, incorporating the DVFS configuration, the thermal behaviour of the data room, and their impact on the energy consumption, to optimise the integrated management of computing and cooling systems.

Through close synthesis of theory and practice, the proposed approach will deliver timely innovations in fundamental science and translational outcomes with significant economic, environmental, and societal benefits. This ambitious research will address a critical energy challenge facing Cloud computing through systematic alignment of energy efficiency and sustainability. To realise our proposed vision, we discussed various research issues (earlier in Sections 4 and 5) and approaches for addressing as future directions, which can be summarised as the following guidelines:

(1) Create a novel architectural framework and principles for energy-efficient and sustainable Cloud computing.
(2) Invent new algorithms for integrated management of all data centre infrastructure resources to minimise energy consumption of both IT resources such as servers and non-IT resources such as cooling systems.
(3) Invent learning-centric dynamic resource provisioning and scheduling algorithms for execution of applications in Cloud environments in an energy-efficient manner while meeting users' QoS requirements.
(4) Proactively balance the state of the Cloud data centre environment for energy-efficient operation by inventing new algorithms and building machine learning models using monitored data collected from the system.
(5) Design innovative mechanisms and policies for thermal and hot spots management for system resiliency.
(6) Minimise the carbon footprint of Cloud services with the invention of new algorithms for distribution of workloads across geographically distributed, green-energy powered edge and Cloud data centre infrastructures.
(7) Develop software systems incorporating energy-efficient resource management algorithms and application demonstrators for deployment in operational edge and Cloud data centres.

The approaches proposed in this paper will provide the vital quantum research leap and scientific foundation necessary for energy efficiency in next generation Cloud computing. This research will create a paradigm shift by: (1) transforming the resource management approaches from "performance-centric" to "energy-centric", "discrete" to "integrated", and "control-centric" to "learning-centric"; and (2) managing two different competing requirements, users QoS expectations and service providers' energy efficiency goals, to solve the global "energy" challenge.

## Acknowledgements

We would like to thank our colleagues Renata Borovica-Gajic, Maria Rodriguez Read, Rami Bahsoon, Satish Srirama, Amanda Jayanetti, Muhammed Tawfiqul, John Grundy, Vlado Stankovski, Rajeev Muralidhar, Adel Nadjaran Toosi, Rodrigo N. Calheiros, Yogesh Sharma, Rao Kotagiri, San Murugesan, Jill Stjohn, and Samodha Pallewatta for their contributions on improving the paper.

This work is partially supported by the University of Melbourne and the HiPEAC6 Network, supported by the European Union's Horizon 2020 research and innovation programme.